\documentclass[twocolumn,showpacs,preprintnumbers,amsmath,amssymb]{revtex4}

\usepackage{graphicx}
\usepackage{dcolumn}
\usepackage{bm}

\begin{document}


\title{Can distributed delays perfectly stabilize dynamical networks?}

\author{Takahiro Omi}
\author{Shigeru Shinomoto}%
\affiliation{%
  Department of Physics, Kyoto University, Sakyo-ku, Kyoto 606-8502, Japan  
}%

\date{\today}

\begin{abstract}

Signal transmission delays tend to destabilize dynamical networks leading to oscillation, but their dispersion contributes oppositely toward stabilization.
We analyze an integro-differential equation that describes the collective dynamics of a neural network with distributed signal delays.
With the gamma distributed delays less dispersed than exponential distribution, the system exhibits reentrant phenomena, in which the stability is once lost but then recovered as the mean delay is increased.
With delays dispersed more highly than exponential, the system never destabilizes.

\end{abstract}


\maketitle


Oscillation is not omnipresent in the brain, but is observed only in limited areas such as the cat visual cortex and the olfactory bulb~\cite{oscillation}.
Extensive synchronization occurs in rather pathological conditions such as parkinsonian tremor or epilepsy~\cite{synchronization}.
How is the collective oscillation avoided in the brain composed of neurons that are prone to synchronize?

Recently it has been revealed that distributed delays in the signal transmission contribute toward stabilization, in neural networks~\cite{gong07,omi07}, ecological systems~\cite{eurich03,eurich05}, control engineering~\cite{control}, biology~\cite{macdonald89} or coupled dynamical systems~\cite{cds}.
In the light of this knowledge, wide distributions of neuronal signal transmission delays reported recently~\cite{delay} bear apparent consistency with the absence of extensive oscillation in the brain.

Conversely, the signal transmission delay itself is known to be a destabilizing factor~\cite{instable}.
We therefore wish to comprehend how these factors are competing in real dynamical systems; in particular, how the macroscopic stability of dynamical systems is controlled by the dispersion of delays and their average.

For this purpose, we examine the stability of an integro-differential equation~\cite{macdonald89,delay-equation} derived from microscopic dynamics of a neural network whose signal transmission delays are distributed in time.
It is revealed from the analysis that the network with gamma distributed delays less dispersed than exponential distribution exhibits reentrant stability; the system once destabilizes but then recovers the stability as the average delay is increased. 
With delays dispersed more highly than exponential distribution, the system attains a perfect macroscopic stationarity.

\begin{figure}[h]
\begin{center}
 \includegraphics[width = 80mm]{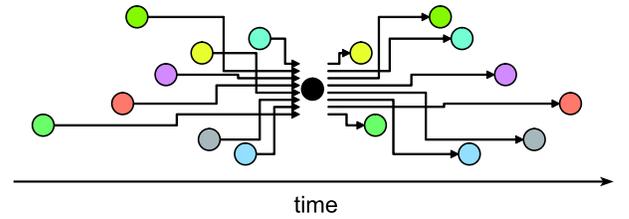}
\caption{
Schematic diagram of distributed signal transmission delays.
}
\label{fig:schematic}
\end{center}
\end{figure}

We consider a network of model neurons that obey the evolution equation,
\begin{equation}
\tau \frac{dx_i(t)}{dt}=-x_i(t)+\textrm{sgn} \left( \sum_{j=1}^{n}w_{i,j}x_{j}(t-d_{i,j})+s_{i} \right),
\label{eq:micro}
\end{equation}
where $\textrm{sgn}(v)$ is the sign function that takes values $+1$ and $-1$ respectively for $v > 0$ and $v \le 0$.
$\tau$ is the ``membrane time constant'' of an individual neuron.
The ``synaptic weight'' $w_{i,j}$ and the signal transmission delay $d_{i,j}$ are fixed to each transmission line from $j$th neuron to $i$th neuron (Fig.\ref{fig:schematic}).
$s_{i}$ is the ``external stimulus'' to each neuron. 

A dynamical equation of the macroscopic order parameter $X(t) \equiv \frac{1}{n}\sum_{i=1}^{n}x_i(t)$ can be derived in a manner similar to what we have done for the discrete time model~\cite{omi07} in parallel with Amari's derivation for the synchronous update rule~\cite{amari71}:
A mean field exerted on $X(t)$ is given by the difference of ratios of positive and negative inputs.
Using the distribution $p_t(v)$ of inputs $\{v_i = \sum_{j=1}^{n}w_{i,j}x_{j}(t-d_{i,j})+s_{i}\}$ at time $t$, the macroscopic equation can be represented as
\begin{equation}
\tau \frac{dX(t)}{dt}=-X(t)+\int_{0}^{\infty} p_t(v) dv - \int_{-\infty}^{0} p_t(v) dv.
\label{eq:pv}
\end{equation}

Under the assumption of the statistical independence of $\{x_j(t-d_{i,j})\}$ from $\{w_{i,j}\}$, the central limit theorem holds for the summed inputs $\{ \sum_{j=1}^{n} w_{i,j}x_j(t-d_{i,j}) \}$.
If in addition $\{s_{i}\}$ are normally distributed, the distribution  $p_t(v)$ can be approximated as Gaussian characterized solely by the mean and the variance at time $t$.
Then, the macroscopic state equation is obtained as 
\begin{eqnarray}
\tau \frac{dX(t)}{dt}=-X(t)+F\left(I\right),
\label{eq:macrofull}
\end{eqnarray}
where $F(x)$ is the error function
\begin{equation}
F(I)=\sqrt{\frac{2}{\pi}}\int_{0}^{I}e^{-\frac{x^2}{2}}dx,
\end{equation}
and $I$ is the ratio of the mean and the standard deviation of inputs $\{v_i\}$ to neurons
\begin{eqnarray}
I = \frac{n \mu_w \langle x \rangle + \mu_s}
{\sqrt{n (\sigma^2_w+\mu^2_w) \langle x^2 \rangle- n \mu^2_w \langle x \rangle^2 + \sigma^2_s}},
\end{eqnarray}
where $\mu_w$, $\mu_s$, $\sigma^2_w$ and $\sigma^2_s$ are respectively the means and variances of $\{w_{i,j}\}$ and $\{s_{j}\}$.

Under the assumption that microscopic states $\{x_i\}$ are statistically independent from delays $\{d_{i,j}\}$, the mean past activity $\langle x \rangle \equiv \frac{1}{n^2} \sum_{i=1}^n \sum_{j=1}^n x_j(t-d_{i,j})$ can be represented by the macroscopic order parameter~\cite{omi07} as
\begin{equation}
\langle x \rangle \equiv \frac{1}{n^2} \sum_{i=1}^n \sum_{j=1}^n x_j(t-d_{i,j}) = \int_{0}^{\infty} g(s) X(t-s) ds,
\end{equation}
where $g(s)$ represents the distribution of delays.
$\langle x^2 \rangle \equiv \frac{1}{n^2} \sum_{i=1}^n \sum_{j=1}^n x^2_j(t-d_{i,j})$ remains close to unity, if individual microscopic states are always approaching swiftly either of $\pm1$.

If $n(\sigma^2_w+\mu^2_w) \ll \sigma^2_s$, or if $\langle x^2 \rangle$ can be approximated as unity, then the dynamical equation (\ref{eq:macrofull}) is closed with the macroscopic order parameter $X(t)$.
Furthermore, if the model parameters satisfy $n\mu_w^2 \ll n\sigma_w^2+\sigma_s^2$, the evolution equation simplifies to
\begin{eqnarray}
\tau \frac{dX(t)}{dt}=-X(t)+F\left(W \int_{0}^{\infty} g(s) X(t-s) ds + S \right),
\label{eq:macro}
\end{eqnarray}
where $W = n\mu_w/\sqrt{n\sigma_w^2+\sigma_s^2}$ and $S = \mu_s/\sqrt{n\sigma_w^2+\sigma_s^2}$.
Note that $n\mu_w^2 \ll n\sigma_w^2+\sigma_s^2$ is not an essential condition for a macroscopic equation (\ref{eq:macro}) to hold but is merely introduced to make the analysis simpler.

With this integro-differential equation, we investigate the influence of the dispersed delays on the macroscopic stability of the system.
We adopt here the gamma distributed delays,
\begin{equation}
 g(s)= \frac{\kappa}{\Gamma(\kappa)T} \left( \frac{\kappa s}{T} \right)^{\kappa-1} \exp{\left(-\frac{\kappa s}{T}\right)},
 \label{eq:gamma}
\end{equation}
which is characterized by the scale factor $T$ representing the average delay, and the shape factor $\kappa$ representing (inversely related to) the dispersion of delays.
$\Gamma(\kappa)$ is the gamma function defined by $\int_0^{\infty} x^{\kappa-1} e^{-x} dx$.

\begin{figure}[h]
\begin{center}
 \includegraphics[width = 80mm]{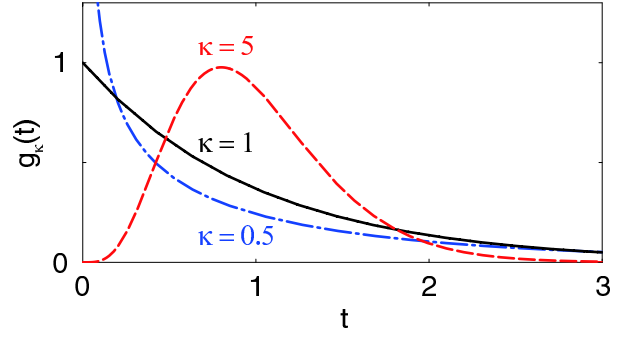}
\caption{
The gamma distributions of identical means ($T=1$) with different shape factors; Exponentially distributed ($\kappa=1$), with the standard deviation identical to the mean;
More dispersed ($\kappa<1$), with the standard deviation greater than the mean;
Less dispersed ($\kappa>1$), with the standard deviation smaller than the mean.
}
\label{fig:gamma}
\end{center}
\end{figure}


Given a macroscopic stationary state $X(t)=X_0$ that satisfies $X_0=F\left(WX_0+S\right)$, we examine the linear stability with respect to the deviation from the stationary state, by putting $X(t)-X_0 = \exp(\lambda t/\tau)$.
The characteristic equation for the linearized integro-differential equation is obtained as
\begin{equation}
(1+\lambda)(1+\lambda T / \tau \kappa)^\kappa=\beta,
\label{eq:character}
\end{equation}
where $\beta$ represents the slope of the response function,
\begin{equation}
\beta \equiv \left. \frac{dF(WX+S)}{dX}\right|_{X=X_0}.
\label{eq:beta}
\end{equation}

For positive $\beta$, the system exhibits instability, $\lambda \ge 0$, if the response function $F(WX+S)$ has a slope greater than unity, $\beta>1$.
In this case, the system eventually attains one of stable stationary states due to the nonlinear saturation.

A dynamical instability leading to oscillation may take place for negative $\beta$.
In this case, the linear-stability boundary is obtained by solving a pair of simultaneous equations that represent the condition for the characteristic equation (\ref{eq:character}) to have a pure imaginary solution $\lambda=i \omega$:
\begin{eqnarray}
\arctan(\omega)+\kappa \arctan(T \omega/\tau \kappa)= \pi,\label{eq:omega}\\
\beta^2 = (1+\omega^2) (1+(T \omega/\tau \kappa)^2)^{\kappa}.\label{eq:beta}
\end{eqnarray}

Equation (\ref{eq:omega}) has a solution only if $\kappa>1$.
In the limit of $T/\tau \to 0$, $\omega$ is obtained as
\begin{equation}
\omega \approx (\kappa \tau/T) \tan \left( \pi/2\kappa \right),
\end{equation}
and critical $\beta$ is obtained from Eq.(\ref{eq:beta}) as
\begin{equation}
\beta \approx -(\kappa \tau/T) \tan(\pi/2\kappa)(1+\tan^2(\pi/2\kappa))^{\kappa/2}.
\end{equation}
In the opposite extreme of $T/\tau \to \infty$, $\omega$ is obtained as
\begin{eqnarray}
\omega &\approx& (\kappa \tau/T) \tan \left( \pi/\kappa \right), \,\, (\kappa > 2),\\
\omega &\approx& \tan \left( \pi (1-\kappa/2) \right), \,\, (2 > \kappa > 1),
\end{eqnarray}
with which the critical $\beta$ is obtained respectively as
\begin{eqnarray}
\beta &\approx& -(1+\tan^2(\pi/\kappa))^{\kappa/2} , \,\, (\kappa > 2),\\
\beta &\approx& -\{(T/\tau \kappa)\tan(\pi(1-\kappa/2))\}^{\kappa/2}\nonumber \\
&\times& \{1+\tan^2(\pi(1-\kappa/2))\}^{1/2}, \,\, (2 > \kappa > 1).
\end{eqnarray}

\begin{figure}[h]
\begin{center}
 \includegraphics[width = 80mm]{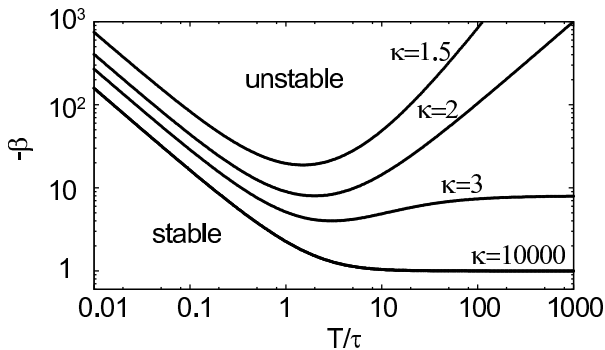}
\caption{
The stability boundaries in the space of $\{T/\tau, -\beta\}$ obtained for several shape parameters, $\kappa=1.5$, $2$, $3$, $10000$.
}
\label{fig:phase}
\end{center}
\end{figure}

Figure \ref{fig:phase} depicts the stability boundaries in the phase space of $\{T/\tau, -\beta\}$ obtained by numerically solving Eqs.(\ref{eq:omega}) and (\ref{eq:beta}) for several shape parameters $\kappa$:
With identical delays ($\kappa \to \infty$), the critical coupling strength $|\beta|$ decreases monotonically with the average delay $T$.
In other words, the system becomes more fragile with delays.
This fact is consistent with the knowledge that the transmission delay is a destabilizing factor.
It is notable that a system with dispersed delays exhibits reentrant phenomena; the stability is once lost but then recovered as the average delay $T$ is increased:
With the delays of small dispersion ($2 < \kappa < \infty$), the critical $|\beta|$ rebounds and then saturates to a finite value.
In a middle range of dispersion ($1 < \kappa \le 2$), the critical $|\beta|$ takes a minimum and diverges with $T$.
With the delays highly dispersed ($\kappa \le 1$), the network never destabilizes.


Next, we numerically solve the macroscopic evolution equation (\ref{eq:macro}) to see if there is nontrivially coexisting oscillation or chaos in the parameter range in which the linear stability is confirmed.
For this purpose, we have tested a number of random initial conditions of various power spectra; from white (jagged) to colored (smooth) random temporal patterns.

Solving an integro-differential equation is generally a hard computational task.
In some particular conditions, however, the computational complexity can be reduced drastically by devising efficient algorithms:
For the exponential distribution of delays ($\kappa=1$), the mean past activity $\int_{0}^{\infty} e^{-s} X(t-s) ds$ is represented by
\begin{equation}
A_t=\sum_{j=0}^{\infty} e^{-j \Delta} X(t-j \Delta),
\end{equation}
where $\Delta$ is a unit step.
Due to the exponential kernel, $A_t$ can be obtained by simply iterating the recurrence equation:
\begin{equation}
A_t= e^{-\Delta} A_{t-\Delta} + X(t) \times \Delta.
\end{equation}

In addition, for the case of $\kappa=2$, the computational complexity could be reduced by utilizing the relation of
\begin{equation}
t e^{-t} \approx \frac{e^{-t}-e^{-(1+\varepsilon)t}}{\varepsilon},
\end{equation}
with sufficiently small $|\varepsilon|$.

\begin{figure}[h]
\begin{center}
 \includegraphics[width = 80mm]{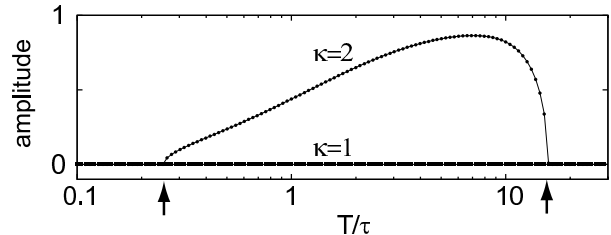}
\caption{
The amplitude of $X(t)$ numerically obtained for the cases of $W=-25$ and $S=0$, ($\beta=-20$);
The linear instability boundaries for the case of $\kappa=2$ are $T/\tau=0.254$ and $15.7$ and depicted by the arrows. 
They coincide with the critical points at which the system numerically shows reentrant stability;
The system with delays exponentially distributed ($\kappa=1$) always remains stable.
}
\label{fig:bifurcation}
\end{center}
\end{figure}

To the extent we have exhausted, we have not found any non-trivial coexisting dynamical orbits in the parameter range that the linear stability is guaranteed.
Figure \ref{fig:bifurcation} shows an amplitude of $X(t)$ obtained for the systems with shape parameters of $\kappa=2$ and $1$.
In the case of $\kappa=2$, a significant amplitude of $X(t)$ can be observed in the interval of mean delay $T$ in which the system is linearly unstable.
For the shape parameter $\kappa=1$ with which the system is linearly stable, the amplitude is negligibly small in the whole range of $T$.

What is the key factor in the perfect stability?
As the system shows a perfect stability in the absence of delay, we suspect if a fraction of instantaneous signal transmissions lead to the stability for $\kappa \le 1$.
We examine the stability of a system with the delay distribution composed of two delta functions peaked at zero and finite delays, $a \delta(t)+(1-a) \delta(t-T)$.
It is found from the characteristic equation that the system can be destabilized even if delay-less lines are present in a finite fraction, $0 \le a < 1/2$.
This fact demonstrates that the presence of instantaneous signal transmissions alone does not necessarily induce a perfect stability.

\begin{figure}[h]
\begin{center}
 \includegraphics[width = 80mm]{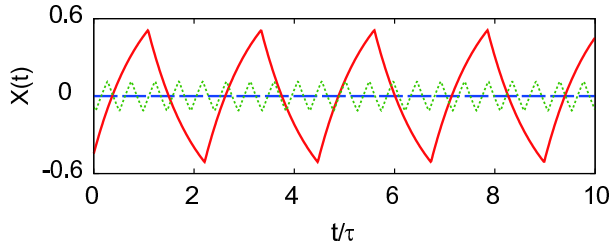}
\caption{
The macroscopic order parameters $X(t)$ numerically obtained for the cases of $W=-1250$ and $S=0$, ($\beta=-1000$), and $T/\tau=1$;
Solid line: Oscillation observed for the shape parameter $\kappa=2$;
Dashed line: Stability observed for $\kappa=1$;
Dotted line: Rapid oscillation of small amplitude observed for the gamma distributed delays of $\kappa=1$ accompanied by a time lag of $\epsilon/\tau=0.01$.
}
\label{fig:dynamics}
\end{center}
\end{figure}

Next, we suspect if the long tail of the delay distribution has led to the stability.
We examine whether or not a system remains stable even if a lag is added to gamma distributed delays (\ref{eq:gamma}) as $\theta(t-\epsilon) g(t-\epsilon)$, where $\theta(x)$ is the Heaviside step function: $\theta(x)=1$ if $x \ge 0$ and $=0$ otherwise.
In the presence of a lag, $\epsilon>0$, the instability condition Eq.(\ref{eq:omega}) is modified as
\begin{eqnarray}
\arctan(\omega)+\kappa \arctan(T \omega/\tau \kappa)+\epsilon \omega/\tau = \pi.\label{eq:omegalagg}
\end{eqnarray}
This characteristic equation possesses a solution even for the case of a high dispersion, $\kappa \le 1$.
This implies that the perfect stability can be destroyed by a time lag.
As long as the time lag $\epsilon$ is small, however, the system exhibits an instability at a very high frequency of the order of $\tau \pi/\epsilon$, and the amplitude of the order parameter $X(t)$ cannot grow large due to the nonlinearity of the system.
This point is verified by directly solving the original nonlinear macroscopic evolution equation.
Figure \ref{fig:dynamics} compares the order parameters $X(t)$ computed for the several kinds of delay distributions: 
the system of $\kappa=2$ may exhibit a large amplitude $X(t)$; 
the system of $\kappa=1$ has never yielded a significant amplitude $X(t)$; 
the system with gamma distributed delays of $\kappa=1$ accompanied by a small time lag $\epsilon/\tau=0.01$ has yielded an order parameter $X(t)$ rapidly oscillating with a small amplitude.

In the present study, we examined the stability of a neural network whose signal transmission delays are distributed in time.
The network is found to exhibit a reentrant stability for the delays less dispersed than the exponential distribution.
The network attains a perfect stability for the highly dispersed delays.
It is noteworthy that Eurich et al also proved that the perfect stability is attained in the limit of highest dispersion, $\kappa=0$, for an ecological feedback system~\cite{eurich05}.
In this Letter, we have revealed that the perfect stability is manifested in a finite range of the dispersion or the shape parameter, $0 < \kappa \le 1$, for a neural network.
It is desirable to examine the generality of the present findings; whether or not the perfect stability is achieved solely due to the delay distribution, irrespective of detailed dynamics of individual elements.


\section*{Acknowledgments}
This study is supported in part by Grants-in-Aid for Scientific Research to S.S. from MEXT Japan (16300068, 18020015), and by the 21st century COE ``Center for Diversity and Universality in Physics''.


\end{document}